\documentclass[11pt,twoside]{article}

\usepackage{asp2006}
\usepackage{epsf}
\usepackage{psfig}
\usepackage{lscape}
\usepackage{graphicx}

\begin{document}

\title{A VLBI Search for the Origin of Wobbling in Blazar Jets} 

\author{I. Agudo$^{1}$}   

\affil{ $^{1}$Instituto de Astrof\'{\i}sica de Andaluc\'{\i}a (CSIC), Apartado 3004, E-18080 Granada, Spain}    

\begin{abstract} 
An increasing number of blazars have been reported to show jet wobbling (i.e., non-regular rotations of the structural position angle of their innermost jets in the plane of the sky with amplitudes between 20 deg. and 50 deg., and time scales between 2 yr. and 20 yr.).
The physical origin for the observed jet wobbling is still poorly understood, but as this phenomenon is triggered in the innermost regions of the jets, it must be tied to fundamental properties of the inner regions of the accretion system. 
Thus, jet wobbling may be an interesting potential tool for supermassive black hole, accretion and jet launching studies.
As part of a joint theoretical/numerical and observational effort to characterize the observational properties and differences between these three possible scenarios we have started a long-term polarimetric phase-reference 43\,GHz VLBA monitoring program to observe the jet structure and the absolute motion of the jet core of four of the blazars which have shown some of the clearest evidence of large amplitude jet wobbling: NRAO~150, OJ\,287, 3C\,273, and 3C\,345 .
Here we present this project and we argue about its suitability for future VSOP-2 observations. 
\end{abstract}


\section{What is Jet Wobbling?}
\label{whatis}

\begin{figure}
\centering
\includegraphics[width=10cm,clip]{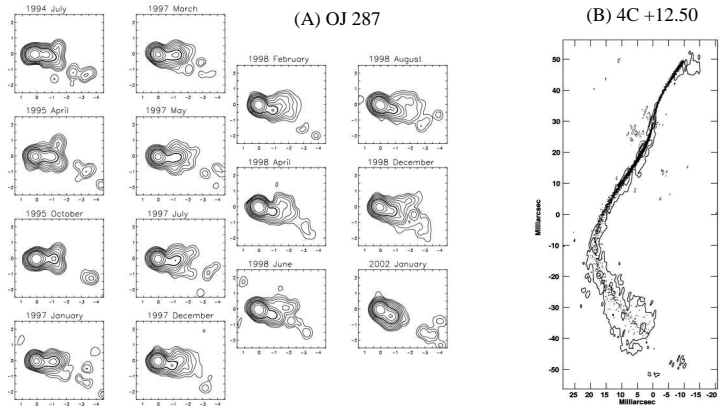}
\caption{{\bf A)} Sequence of 8~GHz VLBA images of OJ~287 from 1994 to 2002. The change of the position angle of the initial jet axis is remarkable even by visual inspection \citep[reproduced from][ with the kind permission of ApJ]{Tat04}. {\bf B)} 15~GHz VLBA image of 4C~$+12.50$. Superimposed is the sky representation of the best-fit streaming model by \citet{Lis03} (with permission by ApJ).}
\label{OJ-4C}
\end{figure}

\begin{figure}
\centering
\includegraphics[width=13cm,clip]{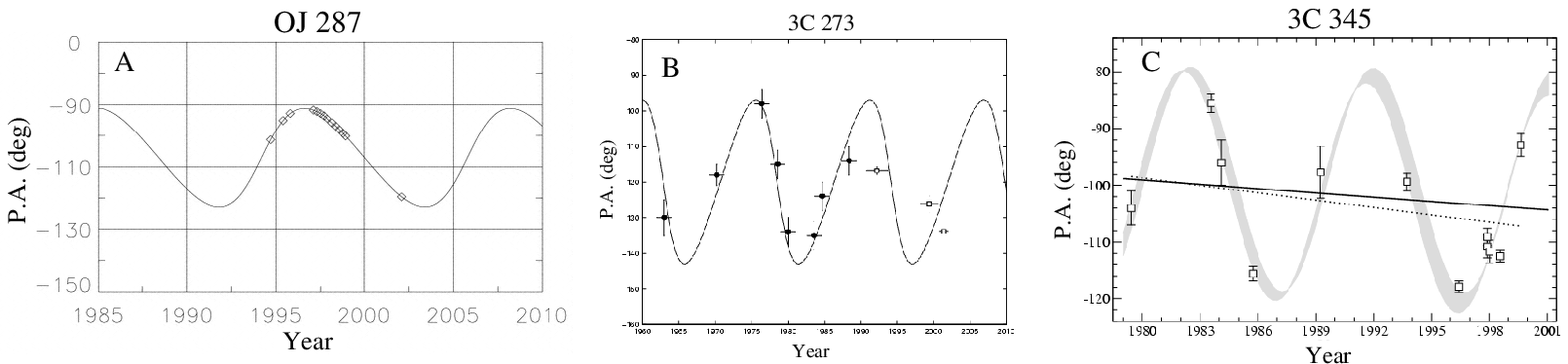}
\caption{Jet structural position angles of OJ\,278 \citep[A; from][with ApJ permission]{Tat04}, 3C\,273 \citep[B; from][with the kind permission of A\&A]{Sav06}, 3C\,345 \citep[C; from][with A\&A permission]{Lob05}.}
\label{PAvsT}
\end{figure}

High resolution VLBI observations are revealing an increasing number of jets in AGN (in particular in blazars) showing either regular or irregular swings of the innermost jet structural position angle in the plane of the sky \citep[e.g. see Fig.~\ref{OJ-4C}, and][]{Sti03,Bac05,Lob05,Sav06,Agu07}.
Time scales between 2 and 20 years and structural position-angle oscillations (projected in the plane of the sky) with amplitudes from $\sim 25^{\circ}$ to $\sim 50^{\circ}$ are typical for the reported cases, e.g. Fig.~\ref{PAvsT}.
This phenomenon, which is detected only in the innermost regions of jets in blazars (essentially within the parsec scale), will be called hereafter \emph{jet wobbling}.
Parsec scale AGN jet curvatures and helical-like structures observed in the inner parsecs and at larger distances from the central engine are also believed to be triggered by changes in direction at the jet ejection nozzle \citep[e.g. Fig.~\ref{OJ-4C}, and][]{Dha98}.

\section{Jet Wobbling in NRAO~150}
\label{nrao}

\begin{figure}
\centering
\includegraphics[width=12cm,clip]{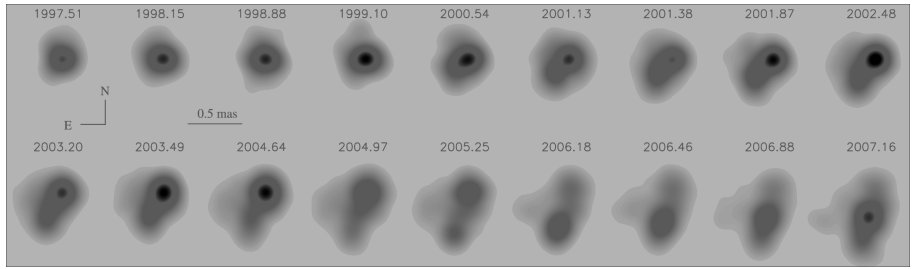}
\caption{Selection of 18 NRAO~150 images obtained from 1997 to 2007 with the VLBA  at 43~GHz \citep[][]{Agu07}. 
All images have been convolved with the same circular beam of 0.16~mas. 
The common intensity scale ranges from 0.025~Jy/beam to 4.128~Jy/beam.
A movie corresponding to these images can be downloaded from www.aanda.org/articles/aa/full/2007/48/aa8448$-$07/movie$\_$AA$\_$476$\_$3$\_$L17.avi}
\label{NRAO150}
\end{figure}

Recently, the discovery of an extreme case of jet wobbling in the quasar NRAO~150 has been reported through the first ultra-high-resolution VLBI set of images obtained from this source at 86~GHz (3.5~mm) and 43~GHz (7~mm) \citep[see Fig~\ref{NRAO150}][]{Agu07}.
To compute the first quantitative estimates of the basic physical properties of the jet in NRAO~150, we have analyzed the ultra-high-resolution images from a new sub-milliarcsecond-scale data set  (covering from 1997 to 2007) at 86~GHz and 43~GHz with the GMVA, the Global Millimetre VLBI Array (GMVA, see www.mpifr$-$bonn.mpg.de/div/vlbi/globalmm/) and the VLBA.

The data shows an extreme projected counter-clock-wise jet wobbling of up to $\sim11^{\circ}/\rm{yr}$ within the inner $\sim61$~pc of the jet, which is associated with a highly {\it non-ballistic superluminal motion} of the jet within this region.
It is worth stressing that the non-radial components of the projected velocity vector with respect to the core position of the main jet features range between 2.7~$c$ and 1.4~$c$.
This may sound surprising because, in AGN, superluminal motions are usually reported in the radial direction relative to the core only.
Recently however, a growing number of AGN jets exhibiting highly bent model-component trajectories have been reported \citep[e.g.,][]{Hom01,Jor05}.
NRAO~150 seems to be an extreme case, which reflects the remarkable non-ballistic nature of its jet features.

In view of this intriguing result, we argue that the magnetic field might play an important role in the dynamics of the jet in NRAO~150, which is supported by the large values of the magnetic field strength obtained from our first estimates \citep[$B\approx0.7$\,G.][]{Agu07} .
It is still unclear whether the reported change of the direction of ejection in {NRAO~150} is related to a regular (strictly periodic or not) behavior or to a single event.
In any case, the extreme characteristics of the observed jet wobbling make {NRAO~150} a prime source to study the jet wobbling phenomenon, although its wobbling time scale ($>15-20$\,yr.) seems to be larger than other well known cases (see Section~\ref{whatis}).

\section{Scientific Interest of Jet-Wobbling VLBI Studies}
\label{motiv}

As outlined in Section~\ref{whatis}, the jet wobbling phenomenon is only observed in the innermost regions of the jets in AGN.
This implies that whatever it is the cause, it should be related to the inermost properties of the jet and perhaps also with the properties of the accretion system, which also implies the properties of the supermassive black hole.
This makes jet wobbling an interesting potential tool for supermassive black hole, accretion and jet launching studies.

\begin{figure}
\centering
\includegraphics[width=11cm,clip]{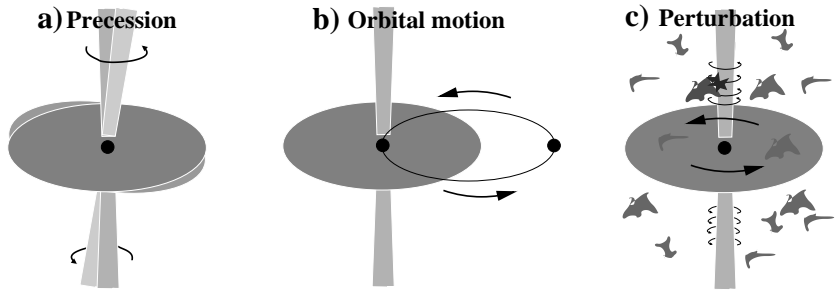}
\caption{Conceptual illustration of three of the possible scenarios causing blazar jet wobbling.}
\label{scenario}
\end{figure}

However, the physical origin for the observed jet wobbling is still poorly understood. 
This phenomenon could be triggered, in principle,  by different causes (see Fig.~\ref{scenario} for a sketch):
{\bf a)} accretion--disk precession,
{\bf b)} orbital motion of the accretion system or
{\bf c)} some other kind of more erratic disk or jet instabilities (e.g. similar to those thought to produce the quasi periodic oscillations [QPO] in X-ray binaries or perturbed rotation of the jet flow around its axis).
There is still no general paradigm to explain the phenomenon of jet wobbling, however, disk precession (cause {\bf a)}) seems to be nowadays the preferred assumed scenario to test and model the quasi-regular jet structural position and integrated emission variability of AGN \citep[e.g.,][]{Cap04}.
Most AGN precession models are driven either by a companion super-massive black hole or another massive object which induces torques in the accretion disk of the primary \citep[e.g.,][]{Lis03,Sti03}, or by the Bardeen-Peterson effect \citep[e.g.,][]{Liu02}.
However, the orbital motion of the jet nozzles (cause {\bf b)} above), although it has some problems regarding the wobbling time scales \citep[see][]{Lob05}, can not be completely ruled yet.
Cause {\bf b)} has not still been extensively considered for the case of jets in AGN, but it might apply as well.

\begin{figure}
\centering
\includegraphics[width=11cm,clip]{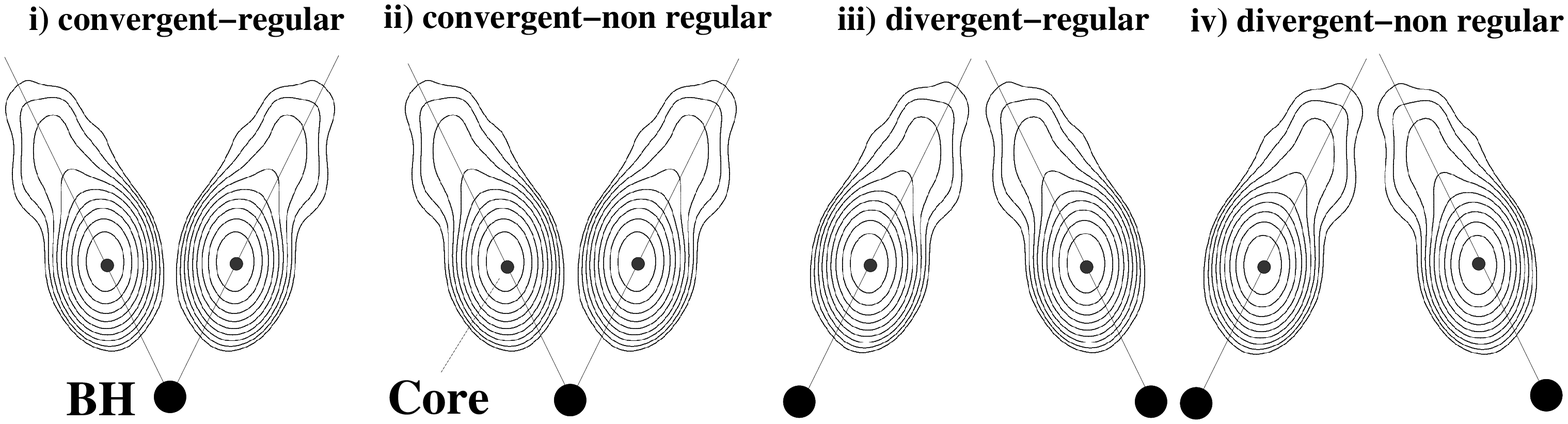}
\caption{Conceptual illustration of the four observational scenarios. A fifth one (with the BH located at the position of the core of the source due to projection effects and to a short distance from the BH to the unprojected core) has not been considered here.}
\label{obscenario}
\end{figure}

\section{Observational Considerations}
\label{cons}

Hence, aiming at looking for the origin of jet wobbling and to extract information about the accretion system and the innermost regions of the jets (including the properties of the SMBH, the accretion disk, and the jet launching region), we have started a new observational (VLBI) and numerical (involving RMHD and synchrotron emission simulations, and jet perturbation theory, see Perucho's paper in this conference proceedings) program to characterize the properties of different scenarios.

The numerical approach is essential to disentangle between the different phenomenologies related to the  three proposed scenarios (Section\ref{motiv}), which are influenced by non-linear processes inherent to the RMHD flows. 
The magnetic field might have a relevant influence as suggested by the results from the NRAO150 observations presented in Section~\ref{nrao}

From the observational point of view, an important issue has not been taken into account up to now to approach the problem.
Most of the previous VLBI data analysis assume that the jets wobble with respect to their cores, which are also assumed to have a fixed position on the sky.
This does not necessarily have to be true. 
In fact, \citet{Mar02} estimated a distance of the jet core to the central engine of the AGN in 3C\,120 of $\sim0.3$\,pc (see Marscher's paper in this conference proceedings and Fig.~\ref{obscenario} for a sketch).
The same should apply to other sources.

Hence, any of the previously mentioned hypothetical jet wobbling scenarios (Fig.~\ref{scenario}) can produce systematic motion of the core in one of the ways illustrated in Fig.~\ref{obscenario}. 
Thus, to attack the problem of the origin of jet wobbling from an observational point of view, the main goal should to measure how the initial jet position angle relates to the absolute position of the jet core.
If any of the observational scenarios outlined in Fig.~\ref{obscenario} is identified, this would help to distinguish between different causes for wobbling scenarios through comparison with the results from adequate numerical simulations.

Even VLBI observations alone could provide important clues about the origin of jet wobbling, e.g. if scenarios {\bf ii)} or {\bf iv)} (in Fig.~\ref{obscenario}) are identified, non-regular disk or jet perturbations would be favored as a promising cause ({\bf c)} in Fig.~\ref{scenario} and Section~\ref{motiv}).
Supporting this possibility, it is still under debate whether the observed jet wobbling is strictly periodic or not \citep[see][for the case of BL\,Lac]{Mut05}. 
In addition, if hypothesis {\bf i)} or {\bf iii)} in Fig.~\ref{obscenario} apply, we might have the chance to obtain independent geometrical constraints on the distance from the mm-core to the central compact object, which may provide relevant information for jet launching models.

\section{The New VLBI Program and VSOP-2}
\label{resul}

At present, VLBI observations at millimetre wavelengths are a powerful technique to image the innermost regions of AGN jets -which are self-absorbed at longer wavelengths, see Marscher's paper in this conference proceedings- with the highest angular resolutions;$\sim 0.15$\,mas at 7~mm (43~GHz) and $\sim 50\,\mu$as at 3.5~mm (86~GHz) with ground arrays.
In addition, the largest astrometric precision is needed to measure the absolute position of the core, which is also achieved through mm VLBI.

\begin{figure}
\centering
\includegraphics[width=13cm,clip]{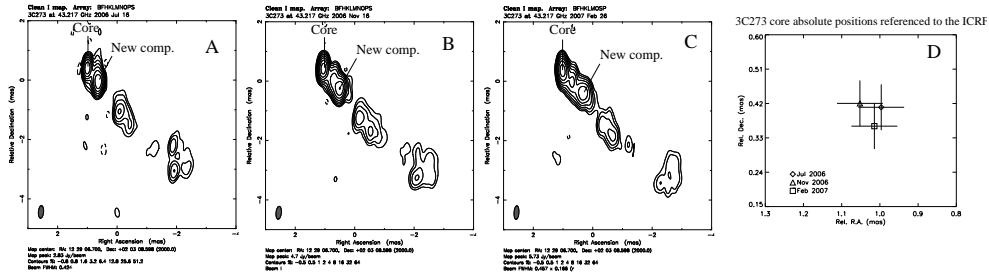}
\caption{43\,GHz VLBA images of 3C\,273 taken on 2006 Jul. (A),  2006 Nov. (B), and 2007 Feb. (C). On the right plot (D) the absolute position of the 3C\,273 core on such epochs, and computed from our phase reference experiment, is presented. Typical 43\,GHz errors for such positions -of 60\,$\mu$as \citep[][]{Gui00}- are symbolized by crosses. It is remarkable the good level of consistency between epochs, which is expected from successful astrometric experiments within time scales lower than 1 yr in the absence of component blending at the core due to new ejections.}
\label{3C273}
\end{figure}

Hence, in May 2006, we started two  43\,GHz VLBA phase reference programs to monitor the absolute motion of the cores and the jet wobbling in NRAO~150, OJ\,287, 3C\,273 and 3C\,345. 
We observe every 6 to 8 months using ``inverse" phase reference (to determine the phase solutions on the strong target sources and transferring those to the position-reference calibrators), switching between two position-reference calibrators, with cycle times $<$120\,s, and with position-reference sources at angular distances between $~\sim0.5^{\circ}$ and at$~\sim3.4^{\circ}$.
43\,GHz VLBA phase reference observations was already demonstrated to be feasible even for sources at $5^{\circ}$ \citep[see][]{Gui00}.
Indeed, that is also shown by the preliminary results from our program (Fig.~\ref{3C273}).

As at present it is not possible, ``a priori", to know the magnitude of the core absolute motion due to the jet wobbling phenomenon, an adequate VLBI study like the one presented here requires the highest possible astrometric precision, which is provided by mm VLBI. Due to the lack of enough sensitivity and to the short coherence time at higher frequencies, in practice 86\,GHz VLBI astrometry is still a challenge for a monitoring program of several sources with reference calibrators at several degrees.
Clearly, 43\,GHz VLBI is more suitable. 
Future space VLBI observations will provide similar resolution and astrometric precision than ground VLBI at 86\,GHz, and about the same sensitivity than the VLBA at 43\,GHz. 
Hence, when operative, VSOP-2 will provide the best tool for observational VLBI jet wobbling studies.

\acknowledgements
The author acknowledges A. Roy., M. Perucho, J. L. G\'omez, A Lobanov, A. Marscher, 
S. Jorstad, M. Roca-Sogorb, J. M. Mart\'i, T. P. Krichbaum, and U. Bach for heir collaboration in the programs outlined in Section~\ref{resul}
The LOC and SOC of this conference are also gratefully acknowledge for their care in the organization of this excellent symposium.
The author has been supported by an I3P contract by the Spanish ``Consejo Superior de Investigaciones Cient\'{i}ficas".
The VLBA is an instrument of the NRAO, a facility of the National Science Foundation of the U.S.A. operated under cooperative agreement by Associated Universities, Inc. (U.S.A.).



\begin{thebibliography}{}

  \bibitem[Agudo et al. (2007)]{Agu07} Agudo, I., Bach, U., Krichbaum T.~P. et al. 2007, A\&A, 476, L17

  \bibitem[Bach et al. (2005)]{Bac05} Bach, U., Krichbaum, T.~P., Ros, E. et al 2005, 433, 815
  
  \bibitem[Caproni et al. (2004)]{Cap04} Caproni, A., Mosquera-Cuesta, H.~J. \&  Abraham, Z. 2004, ApJ, 616, L99

  \bibitem[Dhawan et al. (1998)]{Dha98}  Dhawan, V., Kellerman, K.~I. \& Romney, J. D.  1998, ApJ, 498, L111

  \bibitem[Guirado et al. (2000)]{Gui00} Guirado, J.~C., Marcaide, J.~M. , P\'erez-Torres, M.~A., Ros, E.  2000, A\&A, 353, L37

  \bibitem[Homan et al. (2001)]{Hom01} Homan, D.~C., Ojha, R., Wardle, J.~F.~C.  et al. 2001, ApJ, 549, 840

  \bibitem[Jorstad et al. (2005)]{Jor05} Jorstad, S.~G., Marscher, A.~P., Lister, M.~L.  et al. 2005, AJ, 130, 1418

  \bibitem[Lister et al. (2003)]{Lis03} Lister, M.~L., Kellermann, K.~I.,  Vermeulen, R.~C. et al. 2003, ApJ, 584, 135

  \bibitem[Liu \& Melia (2002)]{Liu02} Liu, S. \& Melia, F. 2002, ApJ, 573, L23

  \bibitem[Lobanov \& Roland (2005)]{Lob05} Lobanov, A.~P. \& Roland, J. 2005,   A\&A, 431, 831

   \bibitem[Marscher et al. (2002)]{Mar02} Marscher, A. P., Jorstad, S.~G., G\'omez, J.~L.  et al. 2002,    Nature,  417, 626

   \bibitem[Mutel \& Denn (2005)]{Mut05} Mutel, R.~L. \& Denn, G.~R. 2005, ApJ,  623, 79

  \bibitem[Pauliny-Toth et al. (1966)]{Pau66} Pauliny-Toth, I.~I.~K., Wade, C. M.  \& Heeschen, D.~S. 1966, ApJS, 13, 65

  \bibitem[Savolainen et al. (2006)]{Sav06} Savolainen, T., Wiik, K., Valtaoja, E. \&  Tornikoski, M. 2006, A\&A, 446, 71

 \bibitem[Stirling et al. (2003)]{Sti03} Stirling, A.~M., Cawthorne, T.~V., Stevens,  J.~A. et al. 2003, MNRAS, 341, 405

 \bibitem[Tateyama \& Kingham (2004)]{Tat04} Tateyama, C.~E. \& Kingham, K.~A. 2004,  ApJ, 608, 149

\end{thebibliography}
\end{document}